\documentclass{aa}  

\usepackage{graphicx}
\usepackage{txfonts}
\newcommand{\sr}{3C~459}
\newcommand{\srr}{3C~459~}
\newcommand{\eg}{{\it e.g.},~}

\begin{document}

   \title{Focusing on the extended X-ray emission in \srr \\ with a {\it Chandra} follow-up observation}

   \titlerunning{X-ray extended emission in \srr from a {\it Chandra} follow-up observation}

   \author{A. Maselli\inst{1,2}
          \and
           R.P. Kraft\inst{2}
          \and
           F. Massaro\inst{3,4,5,6}
          \and
           M.J. Hardcastle\inst{7}}

   \institute{Dipartimento di Fisica, Universit\`a degli Studi di Cagliari, Complesso Universitario di Monserrato, S.P. Monserrato-Sestu km 0.700, I-09042 Monserrato (CA), Italy
         \and
             Harvard - Smithsonian Astrophysical Observatory, Cambridge, MA-02138, USA
         \and
             Dipartimento di Fisica, Universit\`a degli Studi di Torino, via Pietro Giuria 1, I-10125 Torino, Italy
         \and
             Istituto Nazionale di Fisica Nucleare, Sezione di Torino, I-10125 Torino, Italy
         \and
             Istituto Nazionale di Astrofisica, Sezione di Torino, I-10125 Torino, Italy
         \and
             Consorzio Interuniversitario per la Fisica Spaziale (CIFS), via Pietro Giuria 1, I-10125, Torino, Italy
         \and
             School of Physics, Astronomy and Mathematics, University of Hertfordshire, College Lane, Hatfield AL10 9AB, UK
   }

\date{Received 1 May 2018; accepted 25 August 2018}

% \abstract{}{}{}{}{} 
% 5 {} token are mandatory
 
\abstract
% context heading (optional), leave it empty if necessary  
{}
{We investigated the X-ray emission properties of the powerful radio
  galaxy \srr revealed by a recent {\it Chandra} follow-up observation
  carried out in October 2014 with a 62~ks exposure.}
{We performed an X-ray spectral analysis from a few selected regions
  on an image obtained from this observation and also compared the
  X-ray image with a 4.9~GHz VLA radio map available in the
  literature.}
{The dominant contribution comes from the radio core but significant
  X-ray emission is detected at larger angular separations from it,
  surrounding both radio jets and lobes. According to a scenario in
  which the extended X-ray emission is due to a plasma collisionally
  heated by jet-driven shocks and not magnetically dominated, we
  estimated its temperature to be $\sim$0.8~keV. This hot gas cocoon
  could be responsible for the radio depolarization observed in \sr,
  as recently proposed also for 3C~171 and 3C~305. On the other hand,
  our spectral analysis and the presence of an oxygen K~edge,
  blueshifted at 1.23~keV, cannot exclude the possibility that the
  X-ray radiation originating from the inner regions of the radio
  galaxy could be intercepted by some outflow of absorbing material
  intervening along the line of sight, as already found in some BAL
  quasars.}
     {}


   \maketitle

\section{Introduction}

Recent developments in both theory and observations have highlighted
the role of feedback from radio-loud active galactic nuclei (AGNs) in
the evolution of galaxies (\eg
\citealp{1998A&A...331L...1S,2005Natur.433..604D,2013Sci...341.1082M}).
The interaction between radio jets and the interstellar medium (ISM)
inhibits the growth of the supermassive black hole and affects the
star formation rate.
Jets drive large-scale outflows of neutral (\eg
\citealp{2003ApJ...593L..69M}) and ionized (\eg
\citealp{2003ARA&A..41..117C}) gas, which strips away raw materials
needed for star formation.
Moreover, jets would induce turbulence and shocks into the ISM
creating less favorable conditions for the formation of new stars
\citep{2011ApJ...728...29W}.
A considerable amount of gas surrounding jets and lobes, heated and
ionized by shocks in the ISM, is also responsible for the
depolarization observed at radio frequencies (\eg
\citealp{1982ApJ...262..529H,2003MNRAS.339..360H}).
This hot and ionized gas is often associated with the extended
emission line regions (EELR) detected in the optical band and
generally aligned with radio jets (\eg \citealp{1996ApJS..106..281M});
deviations from such an alignment were found in radio galaxies at $z <
0.5$ extending for more than 150~kpc whose emitting line regions show
a filamentary structure extending for a few tens of kiloparsecs
\citep{1989ApJ...336..702B}.

According to this picture, jets are therefore an efficient mechanism
to convert the energy output from the AGN core into an energy input
into the ISM \citep{2012ApJ...747...95G}.
However, detailed descriptions of the shock's behavior have only been
performed for a limited number of nearby sources to date (see, \eg the
analysis carried out by \citealp{2003ApJ...592..129K} for
Centaurus~A).
Therefore, the details of this jet-ISM interaction are not yet firmly
established.
Multifrequency observations of nearby active galaxies are the key to
shedding light on this process.

\begin{figure*}%[tb]
\begin{center}
\includegraphics[width=18.0cm]{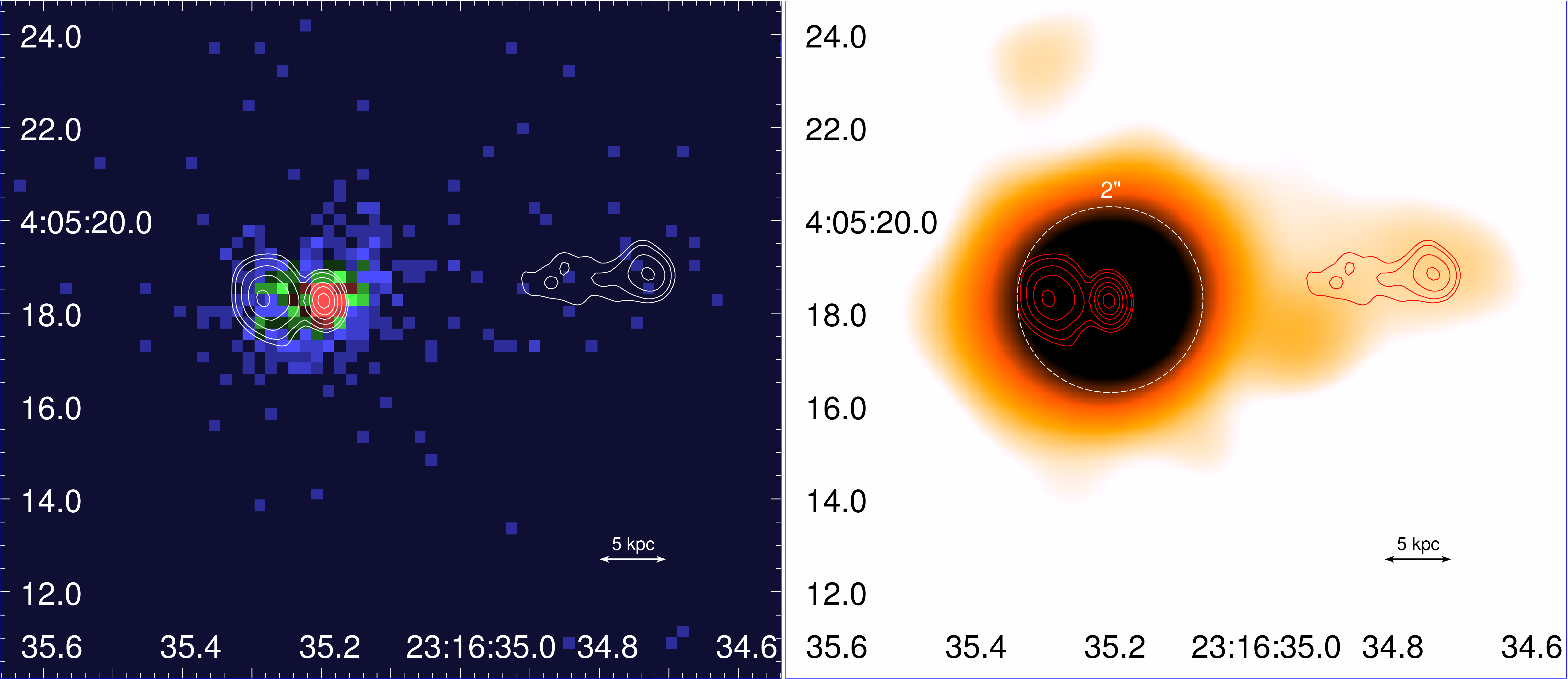}
\caption{{\it Left Panel}: unsmoothed map of the {\it Chandra} count
  rate in the 0.5--7.0 keV band, each pixel corresponding to
  0.246$\arcsec$, that is half the native {\it Chandra} pixel
  size. The radio contours from a VLA map in the C~band have been
  overlaid (white line): they start at 0.48~mJy~beam$^{-1}$ and stop
  at 0.24~Jy~beam$^{-1}$ in correspondence of the core. {\it Right
    Panel}: the {\it Chandra} flux map in the same energy band: a
  circle with a radius of 2$\arcsec$ is centered at the coordinates of
  the core.}
%% no full stop at the end of caption
\label{fig:1}
\end{center}
\end{figure*}

Important contributions have recently come from X-ray observations.
During {\it Chandra} Cycle~9 a snapshot survey was started to complete
the multifrequency database of the revised Third Cambridge catalog
(3CR) with X-ray observations
\citep{2010ApJ...714..589M,2012ApJS..203...31M,2013ApJS..206....7M}.
In addition, recent observations were carried out with the {\it Swift}
satellite to increase the completeness of the sample
\citep{2016MNRAS.460.3829M}.
All 3CR radio sources with $z~\leq~1.5$ have at least an X-ray
snapshot observation present in the {\it Chandra} archive to date
\citep{2015ApJS..220....5M,2018ApJS..234....7M,2018ApJS..235...32S}.
This X-ray database is also enriched by observations that have
comparable angular resolution and performed in the radio, infrared,
and optical bands \citep[see,
  \eg][]{2012ApJ...759...86W,2014ApJ...788...98D,2009A&A...495.1033B,2011A&A...525A..28B,2006ApJS..164..307M,2008ApJS..175..423P,2009ApJS..183..278T,2016ApJS..225...12H}.

The {\it Chandra} snapshot survey of 3CR sources at low redshift
\citep{2010ApJ...714..589M} was also used to plan follow-up
observations of selected targets showing extended X-ray emission
\citep{2012ApJS..203...31M} or with peculiar features, as is the case
for 3C~305 \citep{2009ApJ...692L.123M,2012MNRAS.424.1774H} and 3C~171
\citep{2010MNRAS.401.2697H}.
This is also the case for \sr, for which we present here the results
of our {\it Chandra} follow-up observation.

A brief summary of the multifrequency properties of \srr is presented
in Section~\ref{s:2}, while details of the X-ray data reduction are
reported in Section~\ref{s:3}.
The analysis is then presented in Section~\ref{s:4}, with
Section~\ref{s:5} finally devoted to the discussion of our results and
conclusions.

We assume hereafter a flat cosmology with $H_0 = 72$
km~s$^{-1}$~Mpc$^{-1}$, $\Omega_M~=~0.26$, and
$\Omega_{\Lambda}~=~0.74$ \citep{2009ApJS..180..306D}.
\srr lies at $z~=~0.22$ \citep{2008MNRAS.387..639H} where 1\,arcsec
corresponds to 3.476~kpc and its luminosity distance is 1067.2~Mpc
\citep{2006PASP..118.1711W}.

\begin{figure*}[htb]
\begin{center}
\includegraphics[width=18.0cm]{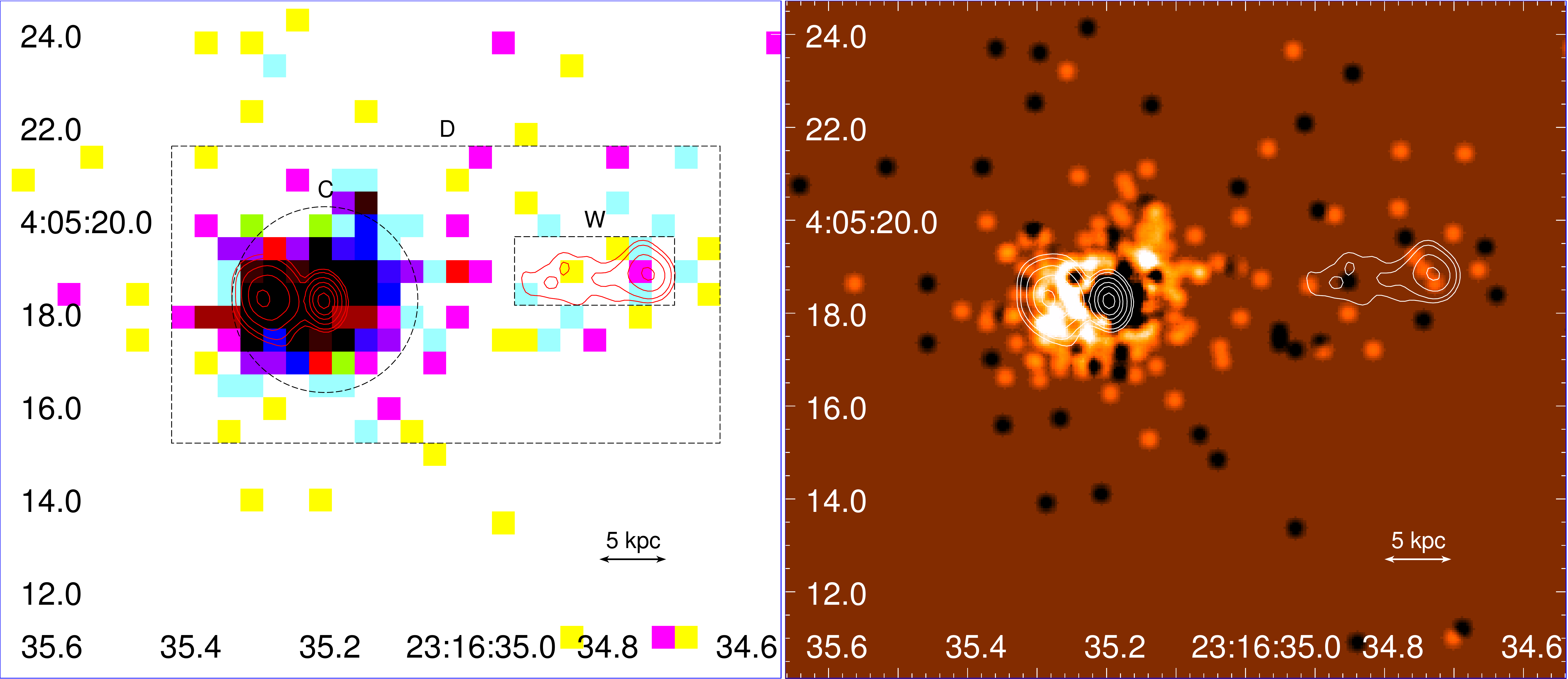}
\caption{{\it Left panel}: Composite RGB image of the X-ray count map
  of \sr. The native pixel size, corresponding to 0.492~arcsec, is
  shown. Different colors distinguish three energy bands: cyan
  (0.5--1.0 keV), magenta (1.0--2.0 keV), and yellow (2.0--7.0
  keV). The contours of adopted extraction regions as described in
  Section~\ref{ss:4.1} are marked with black dashed lines. {\it Right
    panel}: Map of the hardness ratio obtained adopting 0.5--2.0 keV
  and 2.0--7.0 keV as the soft (white points) and the hard (black
  points) X-ray bands, respectively. A smoothing with a Gaussian
  ($\sigma~=~0.123\arcsec$) has been applied.}
% no full stop at the end of caption
\label{fig:2}
\end{center}
\end{figure*}

\section{The peculiar radio galaxy \sr}
\label{s:2}

Source \sr\, is a powerful radio galaxy included in the revised Third
Cambridge Catalog \citep{1985PASP...97..932S} as well as in the
2-Jy~sample \citep{1985MNRAS.216..173W}.
The first detailed radio morphological analysis on this source,
carried out by \cite{1985ApJ...288..514U}, showed a steep-spectrum
core and two edge-brightened lobes with an evident asymmetric profile,
for a total extension of 8.2$\arcsec$ (i.e., 28.5~kpc).
The angular separation between the western lobe and the core is about
five times larger than that from the eastern one (1.3$\arcsec$).
On the basis of its radio morphology, \srr has also been classified as
a compact steep-spectrum (CSS) source (see, \eg
\citealp{1994ApJS...91..491G}; see also \cite{1998PASP..110..493O} for
a review on these objects) though it is not included in the CSS sample
originally assembled by \cite{1989MNRAS.240..657S}.
\cite{2003MNRAS.341...91T} presented a very long baseline
interferometry (VLBI) image of the central core showing a complex
structure consistent with a strongly bent jet, attributed to the
impact of the eastern jet with a large amount of gas in the ISM.
The eastern lobe appears to be significantly depolarized while the
western lobe is not.
This source is thus characterized by a high degree of asymmetry in
both its morphology and depolarization
\citep{1985ApJ...288..514U,1999A&AS..140..355M,2003MNRAS.341...91T}.
Evidence of fast and massive outflows of neutral gas was also reported
by \cite{2005A&A...444L...9M}; \cite{2012ApJ...747...95G} confirmed
this result and traced the presence of ionized gas.

Source \srr has also been intensively observed at higher frequencies.
In the far-infrared, it is about an order of magnitude brighter than
other objects in the 2-Jy sample at similar redshift
\citep{2002MNRAS.330..977T,2009ApJ...694..268D}.

\cite{1981PASP...93..681M} first published an optical-ultraviolet (UV)
spectrum and noticed the presence of the Balmer break and higher
Balmer absorption lines.
This was interpreted as due to photospheres of A-type stars rather
than ISM \citep{1996MNRAS.281..591T}, an interpretation that was later
found to be in agreement with spectral synthesis modeling
\citep{2008MNRAS.385..136W}.
The detection in this spectrum of several emission lines, with no
clear evidence of broad permitted lines, led
\cite{1994ApJ...428...65H} to classify \srr as a narrow-line radio
galaxy.

In \cite{1997MNRAS.286..241J}, \srr was also classified as a high
excitation radio galaxy (HERG; see, \eg \citealp{2012MNRAS.421.1569B})
from the analysis of the optical spectrum published by
\cite{1994ApJS...90....1E}.
This classification was confirmed by \cite{2010A&A...509A...6B} by
computing the value of the excitation index (EI), a spectroscopic
indicator that measures the relative intensity of low and high
excitation lines, using the spectrum published in
\cite{2009A&A...495.1033B}.
In addition, due to the broad shape of the H$\alpha$ line in this
spectrum, \cite{2010A&A...509A...6B} marked \srr as a broad line
object (BLO).
In their search for extended soft X-ray emission in HERG and BLO,
\cite{2012A&A...545A.143B} reported \srr as one of 2 galaxies, among
18 BLOs, showing evidence of such emission and suggested a need for
further investigation.

The host galaxy of \srr shows a disturbed optical morphology, with two
broad symmetrical fan-like protrusions dominated by continuum emission
and extending to the east and the south
\citep{1986ApJ...311..526H,2011MNRAS.410.1550R}.

The information so far collected in the literature for this source is
consistent with a model in which a gas-rich merger episode heated gas
and triggered starburst activity
\citep[see][]{2008MNRAS.385..136W,2011MNRAS.412..960T}.
Subsequently, following the coalescence of the nuclei of the merging
galaxies, the observed radio jets and the overall AGN activity were
also triggered.
\cite{2003MNRAS.341...91T} suggested that the denser gas, responsible
for the bending of the eastern jet, could be the result of this
merging process.

\section{X-ray data reduction}
\label{s:3}

The $\sim$62~ks follow-up observation presented here (ObsID 16044) was
performed on October 12, 2014, with the {\it Chandra} ACIS-S camera
operating in VERY FAINT mode.
The data reduction was carried out following the standard procedures
described in the Chandra Interactive Analysis of Observations (CIAO)
threads~\footnote{(http://cxc.harvard.edu/ciao/guides/index.html)},
and adopted in other analyses (see, \eg
\citealp{2018ApJS..234....7M}).
We used the CIAO software package version~4.8 and the Chandra
Calibration Database (CALDB) version~4.7.2 to process all files.
Level~2 event files were generated using the \textsc{acis process
  events} task, after removing the hot pixels with \textsc{acis run
  hotpix}.
Events were filtered for grades 0, 2, 3, 4, 6 and pixel randomization
was removed.
Astrometric registration was carried out by changing the appropriate
keywords in the fits header so that the nuclear X-ray position was
aligned with the radio one.
The 0.5--7.0~keV count map that we obtained following this procedure
is shown in the left panel of Fig.~\ref{fig:1}.
The same map, although with a different pixel size, is shown in the
right panel of Fig.~\ref{fig:2} emphasizing the different energy of
the detected counts.
For this purpose we distinguished soft (0.5--1.0~keV), medium
(1.0-–2.0~keV), and hard (2.0-–7.0~keV) energy bands.

These same bands were used to create flux maps; a global map in the
0.5--7.0~keV energy range, shown in the right panel of
Fig.~\ref{fig:1}, was also built.
Flux maps were corrected for exposure time and effective area, and our
implementation used monochromatic exposure maps.
We fixed the values for the nominal energies assigned to the soft,
medium, and hard bands to 0.8, 1.4, and 4~keV, respectively; the
exposure maps were built for these nominal values.
Since the natural units of X-ray flux maps are counts
in~cm$^{-2}$~s$^{-1}$, we converted them to cgs units by multiplying
each event by the nominal energy of its band, thereby assuming that
each event in the band has the same energy.
Subsequently, when performing photometry, we applied the correction
factor required to recover the observed units of
erg~cm$^{-2}$~s$^{-1}$.
The ``nominal energy'' was used only to obtain the correct units.
The total energy for any particular region was recovered by applying a
correction factor of E(average)/E(nominal) to the photometric
measurement.
To derive E(average), the actual values were measured with the CIAO
tool {\sc dmstat}.
This correction typically ranged from a few percent to 15\%.
A hardness ratio map in the soft (0.5--2.0 keV) and hard (2.0--7.0
keV) X-ray bands (see Fig.~\ref{fig:2}) was also constructed.

We extracted spectra in the 0.5--7.0 keV energy range using the {\sc
  ciao specextract} script for a few selected regions (see
Section~\ref{s:4} for details).
All spectra were binned to obtain a minimum number of 20 counts per
bin after background subtraction, and were analyzed with {\sc
  xspec}~12.
Uncertainties are reported with a 90\% level of confidence.

\begin{table*}
\caption{The number of counts, and the corresponding surface brightness (counts $\cdot$ arcsec$^{-2}$), computed in different energy bands within a 2$\arcsec$ radius circle $C$ matching the radio core, a rectangle $W$ matching the western radio jet and lobe, the $D$-$C$ region matching the extended emission, and a 50$\arcsec$ radius circle $B$ taking into account the background contribution.} %% no full stop at the end of caption
\label{tab:1}
\centering          
%\begin{center}
\begin{tabular}{c||cc|cc|cc||cc}
%\hline
 Energy band  &     &       $C$        &   &         $W$     &    &      $D$-$C$    &     &        $B$      \\ 
\hline                                                                            
\hline                                                                            
0.5 - 1.0 keV & 146 & 11.599$\pm$0.960 & 3 & 0.590$\pm$0.341 & 29 & 0.461$\pm$0.086 &  69 & 0.009$\pm$0.001 \\ 
1.0 - 2.0 keV & 194 & 15.412$\pm$1.107 & 1 & 0.197$\pm$0.197 & 21 & 0.334$\pm$0.073 & 115 & 0.015$\pm$0.001 \\ 
2.0 - 7.0 keV & 262 & 20.815$\pm$1.286 & 2 & 0.393$\pm$0.278 & 22 & 0.350$\pm$0.075 & 327 & 0.042$\pm$0.002 \\ 
\hline                                                                             
0.5 - 2.0 keV & 340 & 27.011$\pm$1.465 & 4 & 0.787$\pm$0.393 & 50 & 0.794$\pm$0.112 & 184 & 0.023$\pm$0.002 \\ 
0.5 - 7.0 keV & 602 & 47.826$\pm$1.949 & 6 & 1.180$\pm$0.482 & 72 & 1.144$\pm$0.135 & 511 & 0.065$\pm$0.003 \\ 
\end{tabular}
%\end{center}
\end{table*}

\section{X-ray data analysis}
\label{s:4}

\subsection{Imaging analysis}
\label{ss:4.1}
%\subsubsection{}%\label{sss:?}

To carry out the X-ray analysis we selected three regions as described
immediately below and shown in Fig.~\ref{fig:2} (left panel).
These were chosen by overlaying 4.9~GHz radio contours on to the X-ray
image as described in Sect.~\ref{s:3}.

\begin{enumerate}
\item A circle $C$ with a radius of 2$\arcsec$, centered at the
  coordinates used for the astrometric registration and matching the
  core;
\item a rectangle $W$ of size 3.44$\arcsec$ x 1.48$\arcsec$ (21
  pixels) matching both the western jet and lobe;
\item a wider rectangle $D$ of size 11.80$\arcsec$ x 6.40$\arcsec$
  (312 pixels) encompassing the whole radio structure.
\end{enumerate}

The angular separation between a point along the lowest radio contour
and the nearest point on the $D$ perimeter is in the
1.3\arcsec--3.5\arcsec range.
The region resulting from the difference between $D$ and $C$
(hereinafter labeled as $D-C$) was used to study the extended X-ray
emission only.

The detection significance of the X-ray emission within $C$, $W$, and
$D-C$ was evaluated comparing their surface brightness (number of
X-ray counts per unit area) with that computed in a 50$\arcsec$ radius
circle $B$, centered at 90$\arcsec$ east of the core and lacking any
clear X-ray emission due to background sources.
In the 0.5--7.0~keV band the surface brightness in $C$, $W$, and $D-C$
is (47.8~$\pm$~1.9), (1.2~$\pm$~0.5), and
(1.1~$\pm$~0.1)~ct~arcsec$^{-2}$, respectively.
The values computed in $W$ and $D-C$, which are very similar to each
other, are more than an order of magnitude higher than in $B$, thus
resulting in a significant detection.
All details corresponding to this and other energy ranges are reported
in Table~\ref{tab:1}.

\begin{figure}[bht]
\begin{center}
\includegraphics[width=9.6cm,angle=0]{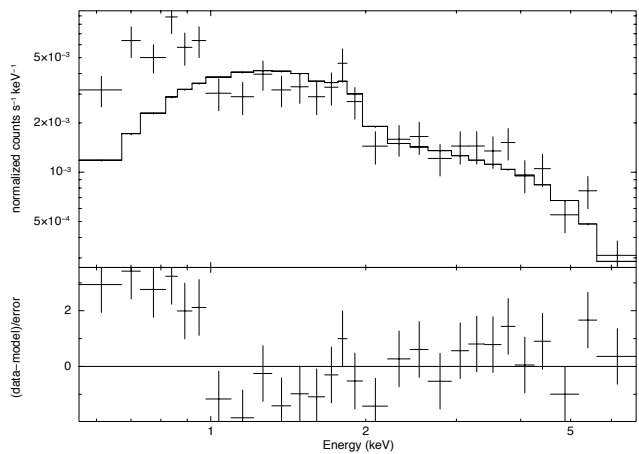}\\
\caption{\footnotesize{Fit to the X-ray spectrum of \sr\, (and corresponding
    residuals) extracted from $C$ in the 0.5--7.0~keV band adopting a
    single, intrinsically absorbed, power law; a soft excess at
    energies lower than 1~keV is evident.}}
% no full stop at the end of caption
\label{fig:3}
\end{center}
\end{figure}

\subsection{Spectral analysis}
\label{ss:4.2}
%\subsubsection{}%\label{sss:?}

For the spectral analysis we assumed that the X-ray emission from \srr
has two main components: the former related to its radio core and the
latter arising from extended emission.
We first describe the analysis of the spectrum extracted from $C$,
where both components are present, and then consider the $D$ spectrum.
All models adopted in our analysis include Galactic absorption with
column density $N_H~=~5.24~\cdot~10^{20}$~cm$^{-2}$
\citep{2005A&A...440..775K}.

\begin{figure*}[h]
\begin{center}
\includegraphics[width=8.7cm,angle=0]{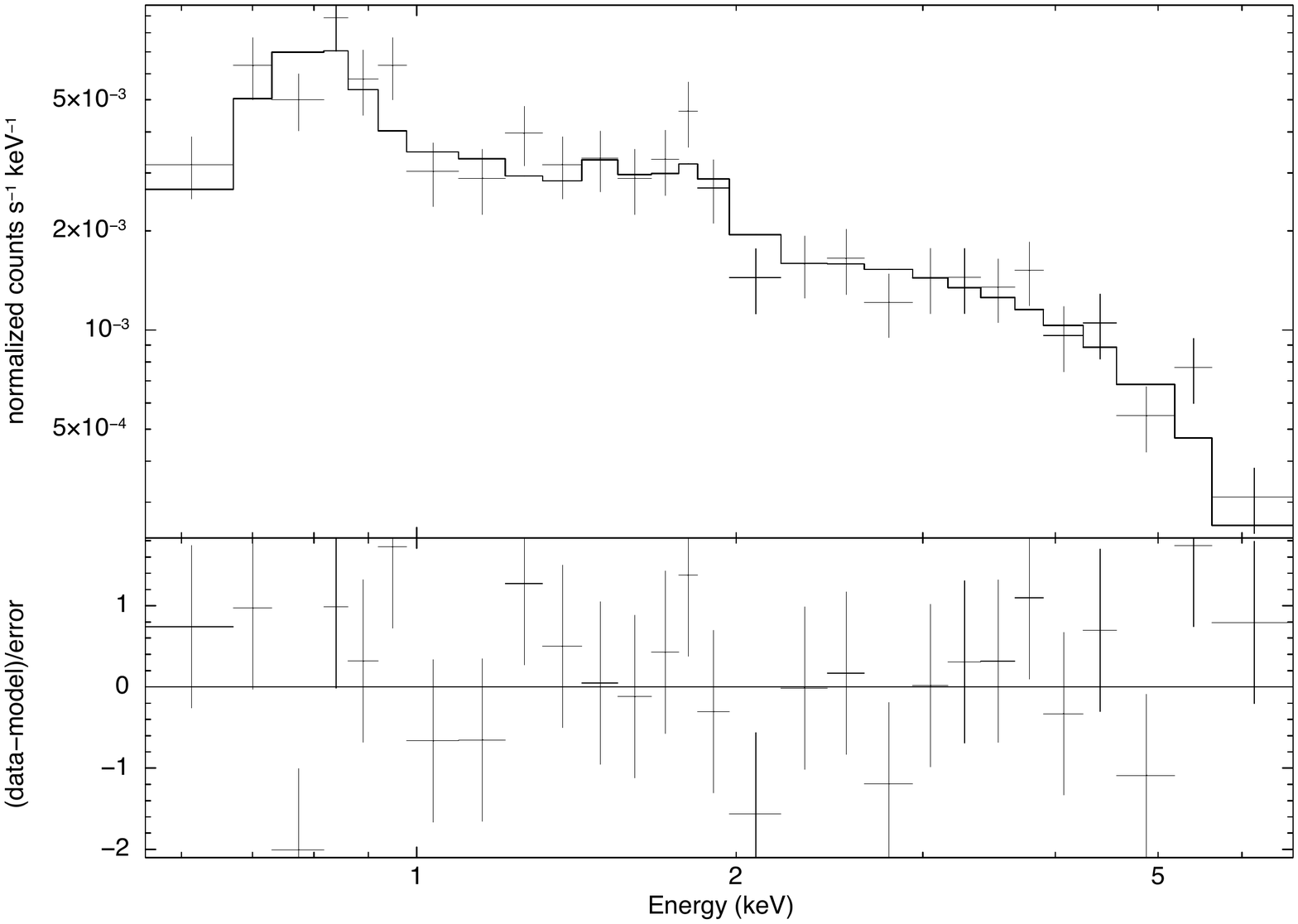}
\includegraphics[width=9.5cm,angle=0]{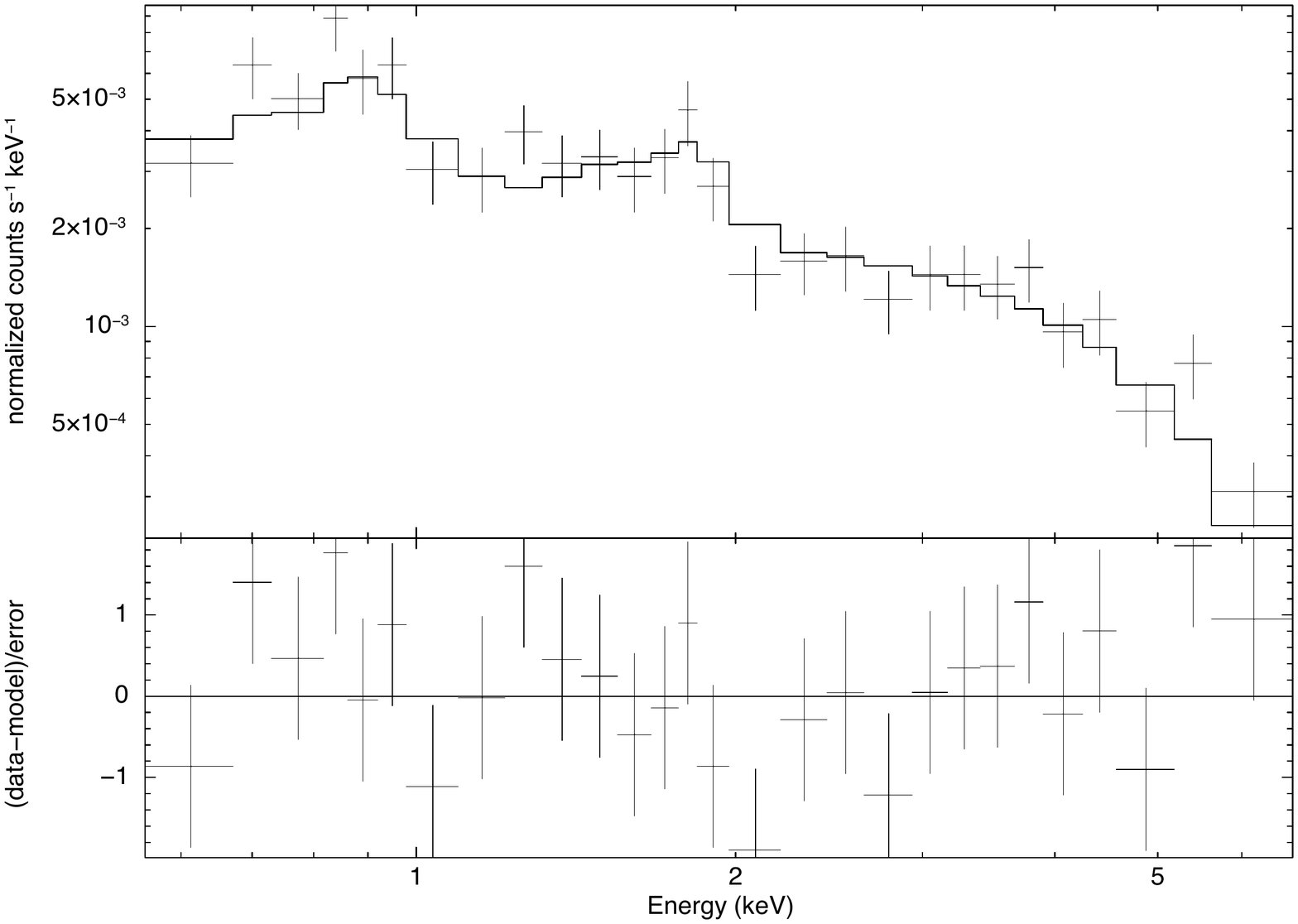}
\caption{{\footnotesize{\it Left panel}: the fit to the X-ray spectrum
    of \sr\, (and corresponding residuals) extracted from $C$ in the
    0.5--7.0~keV band with a model including both the {\sc apec} and
    the absorbed power law. {\it Right panel}: the fit to the same
    spectrum using an {\sc absori} component.}}
% no full stop at the end of caption
\label{fig:4}
\end{center}
\end{figure*}

\subsubsection{X-ray spectrum extracted from $C$}
\label{sss:4.2.1}

The fit of the X-ray spectrum from $C$ with a single unabsorbed power
law model was not acceptable (i.e., $\chi^2$ of 68.33 for 26 d.o.f.).
Fitting with a single, intrinsically absorbed power law did not lead
to any significant improvements.
In both cases we found clear evidence of a soft excess below 1~keV, as
shown in Fig.~\ref{fig:3}.
A fitting procedure with both components improved the $\chi^2$
(i.e., 26.47/23~d.o.f.), but led to a high value of the photon index
$\Gamma_U~=~3.57^{+0.75}_{-0.63}$ for the unabsorbed power law.
We found $N_H~=~(2.91^{+2.35}_{-1.68})~\cdot~10^{22}$~cm$^{-2}$ and
$\Gamma_A=1.39^{+0.59}_{-0.50}$ for the absorbed component: the
$\Gamma_A$ value, though poorly constrained, is consistent within the
errors with $\langle~\Gamma_A~\rangle$~=~1.7 computed considering most
radio-loud AGNs (see, \eg \citealp{2006MNRAS.370.1893H}).

To test the hypothesis that the origin of the soft excess is due to
collitionally ionized plasma, we fitted the spectrum with a model
including both an Astrophysical Plasma Emission Code ({\sc apec}) and
an intrinsically absorbed power law.
Leaving all parameters free to vary we obtained a good description
of the spectrum ($\chi^2$=22.32, 22~d.o.f.) but some parameter values
were poorly constrained: in particular, no improvement was obtained
for $\Gamma_A~=~1.35^{+0.59}_{-0.54}$ while the abundance of elements
in the plasma with respect to solar values was equal to 0.07, and
consistent with zero.
However, such subsolar abundances have already been found in similar
spectral analysis carried out for 3C~171 \citep{2010MNRAS.401.2697H}
and 3C~305 \citep{2012MNRAS.424.1774H}.
Subsequently, we fixed $\Gamma_A~=~1.7$ and the abundance to
0.15~solar, a value that is in line with the results obtained by
\cite{2012MNRAS.424.1774H} (see their Table~3).
In this way we obtained a plasma temperature
$kT~=~0.97^{+0.12}_{-0.21}$~keV and a hydrogen column density
$N_H$~=~(3.37$^{+1.10}_{-0.87}$)$~\cdot~10^{22}$~cm$^{-2}$
($\chi^2~=~25.23$, 24~d.o.f.); this fit and the corresponding
residuals are shown in the left panel of Fig.~\ref{fig:4}.

In light of the distribution of residuals shown in Fig.~\ref{fig:3},
we decided to replace the {\sc zwabs} component in our fitting model
with {\sc zedge}, a redshifted absorption edge.
This component could be interpreted as due to the presence of an
absorbing material intercepting a relativistic outflow along the line
of sight.
As a result, we obtained an acceptable fit ($\chi^2$=27.47, 24~d.o.f.)
with the absorption edge at energy 1.23$^{+0.03}_{-0.04}$~keV and the
photon index at $\Gamma_A$~=~1.55$^{+0.13}_{-0.14}$.
We further replaced an absorption component from ionized material
(i.e., {\sc absori}) with that arising from a neutral absorber (i.e.,
{\sc zedge}).
Following this fitting procedure we found indications of high values
for both the temperature $T_{abs}$ of the absorbing material and the
ionization parameter $\xi$.
Freezing them at $T_{abs}=10^6$~K and $\xi~=~100$, respectively, the
fit improves for increasing values of the Fe~abundance.
Fixing also this last parameter to twice the Solar value, we found
$\chi^2~=~25.25$ (25 d.o.f.), with
$N_H~=~(5.7~\pm~1.3)~\cdot~10^{22}$~cm$^{-2}$ and
$\Gamma_A~=~1.76^{+0.13}_{-0.15}$.
Results of this fit are shown in the right panel of Fig.~\ref{fig:4}.

Finally, we also tested a photoionization scenario to describe the ISM
thermal emission adding up to three emission lines with a Gaussian
profile, with $\sigma$ at the value of 10$^{-3}$~keV, to the simple
power-law component.
Following this procedure, the best result in terms of $\chi^2$ (19.25
with 22 d.o.f.) was obtained including two emission lines at energies
corresponding to the transitions of heavily photoionized elements
(Fe~XVII and Ne~X).
However, the value of the photon index was extremely low
($\Gamma_U~=~0.89~\pm~0.15$).

\subsubsection{X-ray spectrum extracted from $D$}
\label{sss:4.2.2}

Assuming the ({\sc apec} + absorbed power law) description for the $C$
region, we fitted the X-ray spectrum extracted from the $D$ region
with a similar model but fixing the $\Gamma_A~\equiv~1.7$ and $N_H$ to
the value that we obtained from the fit of the $C$ spectrum.
As a result, we found a reliable description of the spectrum
($\chi^2~=~28.08$ with 26~d.o.f.) with the same low value for the
abundance as in the $C$ case, but better constrained
($0.06^{+0.07}_{-0.04}$), and a lower value of the plasma temperature
$kT~=~0.78^{+0.07}_{-0.04}$~keV; the {\sc apec} normalization was
$N~=~(1.10^{+0.69}_{-0.47})~\cdot~10^{-4}$~cm$^{-5}$.

\section{Discussion and conclusions}
\label{s:5}

Here we present results of an X-ray analysis of the follow-up {\it
  Chandra} observation of \srr performed in October~2014 for a total
of 62~ks exposure time.
Our analysis was carried out selecting several regions, as described
in Sect.~\ref{s:4}.
In particular, the $C$ region takes into account the X-ray emission
mainly from the radio core; any eventual contribution from the eastern
lobe could not be disentangled from the core emission due to the small
angular separation (1.3\arcsec) between them.
The $D-C$ region takes into account the extended emission surrounding
the whole radio galaxy: $\sim$11\% of all the counts detected in $D$
correspond to $D-C$.
In this region, the X-ray emission follows the direction traced by the
radio jets, as more clearly visible for the western one (right panel
of Fig.~\ref{fig:1}).
The distribution of the X-ray photons is not confined within the
lowest radio flux density contour, suggesting that they could come
from a plasma that almost uniformly wraps around the whole radio
structure.

In recent years, significant efforts have been made to investigate the
X-ray emission of compact radio galaxies such as CSS and Gigahertz
Peaked Spectrum (GPS)
\citep{2006ApJ...653.1115O,2006A&A...446...87G,2006MNRAS.367..928V,2008ApJ...684..811S,2009A&A...501...89T,2014MNRAS.437.3063K,2016ApJ...823...57S},
and are still ongoing.
Examining the results from a sample of nine bona fide CSS radio
galaxies detected until then in the X rays, \cite{2017ApJ...851...87O}
found convincing evidence for hot shocked gas for just two of them
(3C~303.1 and 3C~305).
However, they noticed that both were not exceptional in their overall
properties, compared to the total sample of nine.
This led them to speculate that hot shocked gas is typical in CSS
sources and would be revealed by performing deeper observations; in
this sense, our results on \srr validate this hypothesis.

In line with a scenario where the extended X-ray emission is due to a
plasma collisionally heated by jet-driven shocks, we estimated a
temperature of 0.78~keV with subsolar abundances.
Assuming a cylindrical geometry with r~=~3.2$\arcsec$ and
h~=~11.8$\arcsec$ for the plasma distribution, which is a
three-dimensional equivalent of the $D$ rectangle, from the {\sc apec}
normalization we derived a hydrogen density
$n_H~=~5.2~\cdot~$10$^{-2}$~cm$^{-3}$.
Under the simple assumption that this plasma is not magnetically
dominated, implying that the energy density in the magnetic field
$B^2$/2$\mu_0$ is lower than the thermal energy density in the gas
(3/2)~$nkT$, this value provides an upper limit for the magnetic field
strength that is $B~\leq$~55~$\mu$G, in agreement with the results
already found for other 3C~sources observed with {\it Chandra} such as
3C~171 \citep{2010MNRAS.401.2697H} and 3C~305
\citep{2012MNRAS.424.1774H}.
The assumption that the gas is not magnetically dominated is supported
by the fact that radio lobes tend to be close to the equipartition and
the magnetic field $B$ is of the order of a few $\mu$G \citep[see,
  \eg][]{2005ApJ...626..733C,2012MNRAS.421..108D}.
Typical $B$ values in radio lobes are similar to $B_{CMB}$, the
magnetic field having equivalent energy density to the cosmic
microwave background (CMB) at the source redshift
\citep{2011ApJ...729L..12M}.
For \srr, $B_{CMB}$ is $\approx$5~$\mu$G, and therefore has an energy
density about one order of magnitude lower than the thermal one.

This hot gas cocoon could be responsible for the radio depolarization
observed in \sr.
Therefore, excluding the contribution from the radio core, whose
emission is expected to be unpolarized, a good spatial match is seen
between the {\it Chandra} image and the radio polarization map (see
middle panel of Fig.~1 in \cite{2003MNRAS.341...91T}).
Only a hint of polarization for the radio emission was found for the
eastern lobe, where a significant number of X-ray counts were detected
(see Fig.~\ref{fig:1}), while a much higher polarization level was
found towards the western lobe, where X-ray emission is fainter.

X-ray emission from the core is considerably absorbed, having a best
fit value of the hydrogen column density
$N_H$~=~(3.37$^{+1.10}_{-0.87}$)$~\cdot~10^{22}$~cm$^{-2}$ obtained
for the absorbed power law component.
This result is also consistent with a scenario, suggested by our
spectral analysis, in which the X-ray radiation, originating from the
inner regions of the radio galaxy, could be intercepted and filtered
from some outflow of absorbing material intervening along the line of
sight.
This possibility has already been discussed by
\cite{2002ApJ...573L..77H} for the broad absorption line (BAL) quasar
APM~08279$+$5255 reporting the detection of an ionized Fe~K~Edge in
the spectrum of an XMM-{\it Newton} observation.
Similar absorbing features in the X-ray spectrum, in the form of lines
rather than edges, have also been detected in a {\it Chandra}
observation on the same source \citep{2002ApJ...579..169C} and in an
XMM-{\it Newton} observation on another BAL quasar, PG~1115$+$080
\citep{2003ApJ...595...85C}.
In the case of \sr, the energy at which we detected the edge is
consistent with a blueshifted oxygen K~edge (0.533~keV); following
\cite{2003A&A...401.1185L} and assuming purely radial motion, we
computed the radial velocity $v_r$ of the plasma needed to produce
this shift in frequency and found $\mid v_r \mid$~=~0.68~c.
Although a redshifted iron K~edge (7.112~keV) cannot be excluded a
priori, it would imply much higher outflow velocities.

Under the assumption that the gas constituting the outflow is ionized,
considerably high values of both the temperature T$_{abs}$ and the
ionization parameter $\xi$ would be requested.
Furthermore, the Fe abundance - higher than the Solar one - estimated
by our spectral analysis could be expected according to the possible
merging event proposed for the \srr evolution and its high star
formation rate.
We note that even higher values, up to approximately five times the
Solar abundance, were reported by \cite{2002ApJ...573L..77H} in their
analysis (see their Fig.~\ref{fig:3}) for a much farther ($z$~=~3.91)
and therefore younger source.

Our analysis confirms that \srr is a very complex and interesting
source that is worthy of further investigation, also supported by new
multi-frequency observations.
In particular, a detailed map in the 5007~\AA\, optical filter,
corresponding to [OIII], with HST would reveal the possible presence
of extended emission line regions (EELRs).
Radiation at this particular frequency is expected to be robust in a
HERG such as \srr (see, \eg \citealp{2010A&A...509A...6B}).
The detection of EELRs would provide useful information on their
distribution with respect to the X-ray emitting regions that we have
revealed, improving our knowledge on the complex multi-phase medium
plausibly permeating the whole radio galaxy.
We also note that, unfortunately, no spectrum at rest-frame
wavelengths shorter than 3500~\AA\, is found in the literature: new
observations in this UV range would be useful to complement the
already rich multi-wavelength information available for this very
peculiar source.

\begin{acknowledgements}

A.M. is grateful to Christine Jones and William Forman for valuable
discussions and the time spent together during his visit at the
Harvard-Smithsonian Astrophysical Observatory. F.M. acknowledges
financial contribution from the agreement ASI-INAF n.2017-14-H.0. This
investigation is supported by the NASA grants GO4-15096X, GO6-17081X
and GO4-15097X. This work is supported by the "Departments of
Excellence 2018 - 2022" Grant awarded by the Italian Ministry of
Education, University and Research (MIUR) (L. 232/2016). This research
has made use of resources provided by the Compagnia di San Paolo for
the grant awarded on the BLENV project (S1618\_L1\_MASF\_01) and by
the Ministry of Education, Universities and Research for the grant
MASF\_FFABR\_17\_01. The authors also thank the anonymous referee for
valuable comments that improved this paper.

\end{acknowledgements}

%\begin{thebibliography}{}

   \bibliographystyle{aa} % style aa.bst
   \bibliography{biblio-u1} % your references Yourfile.bib - join the
% .bib files when you upload your source files
% -------------------------------------------------------------------

%\end{thebibliography}

\end{document}